# Bloch Equation Enables Physics-informed Neural Network in Parametric Magnetic Resonance Imaging

*Qingrui Cai[1,3], Liuhong Zhu[2], Jianjun Zhou[2,3], Chen Qian[3], Di Guo[4], and Xiaobo Qu[1,3]*

[1]National Integrated Circuit Industry Education Integration Innovation Platform, School of Electronic Science and Engineering (National Model Microelectronics College), Xiamen University, Xiamen 361102, China
[2]Department of Radiology, Zhongshan Hospital (Xiamen), Fudan University, Xiamen 361015, China
[3]Department of Electronic Science, Fujian Provincial Key Laboratory of Plasma and Magnetic Resonance, Xiamen University, Xiamen 361102, China
[4]School of Computer and Information Engineering, Fujian Engineering Research Center for Medical Data Mining and Application, Xiamen University of Technology, Xiamen 361024, China.

*Abstract*—Magnetic resonance imaging (MRI) is an important non-invasive imaging method in clinical diagnosis. Beyond the common image structures, parametric imaging can provide the intrinsic tissue property thus could be used in quantitative evaluation. The emerging deep learning approach provides fast and accurate parameter estimation but still encounters the lack of network interpretation and enough training data. Even with a large amount of training data, the mismatch between the training and target data may introduce errors. Here, we propose one way that solely relies on the target scanned data and does not need a pre-defined training database. We provide a proof-of-concept that embeds the physical rule of MRI, the Bloch equation, into the loss of physics-informed neural network (PINN). PINN enables learning the Bloch equation, estimating the $T_2$ parameter, and generating a series of physically synthetic data. Experimental results are conducted on phantom and cardiac imaging to demonstrate its potential in quantitative MRI.

## I. INTRODUCTION

Quantitative magnetic resonance imaging (qMRI) can measure parameters that reflect the intrinsic characteristics of tissues [1]. Common quantitative parameters include $T_1$, $T_2$, $T_2^*$, extracellular volume fraction, etc.. $T_1$, $T_2$, $T_2^*$ and myocardial extracellular volume fraction can be used to detects diffuse fibrosis [3], evaluate the degree of myocardial edema [4], quantifies tissue iron content [5], and reflects the degree of myocardial fibrosis [5], respectively. For example, $T_1$ and $T_2$ parameters (Fig. 1) could assess the histopathological changes of the myocardium and save the myocardium as early as possible [2].

Accurate parameter estimation is important. The estimation include two steps: First, solve the parametric signal model following the physical rule, Bloch equation, of MRI; Second, estimate parameters through fitting the signal model [6]. Thus, parameter estimation may encounter problems if the analytical solution is hard to be obtained [7]. On the other hand, advanced deep learning methods have shown great potential in qMRI [8], but deep learning methods usually lack interpretability and require a large amount of high-quality training labels [9].

Recently, to avoid using the pre-defined training database, physics-informed neural networks (PINN) [10-11] uses known physical equations as prior information and embeds them into the network's loss function.

In this work, we propose a PINN-based method for $T_2$ mapping in this work. We designed a physics-informed loss function that embeds the Bloch equation in PINN. By learning the Bloch equation through the network, it can obtain quantitative $T_2$ values by directly solving the inverse problem of equations without signal analytical formula and with only a single sample data. Once the PINN is trained, it can be used to generate physically synthetic MRI data.

## II. PROPOSED METHOD

### A. Bloch Equation

The magnetization vector $\mathbf{M} = (M_x, M_y, M_z)^T$ in the magnetic field satisfies the Bloch equation:

$$\frac{d\mathbf{M}(\mathbf{r},t)}{dt} = \gamma \mathbf{M}(\mathbf{r},t) \times \mathbf{B}(\mathbf{r},t) - \frac{M_x(\mathbf{r},t)\mathbf{i} + M_y(\mathbf{r},t)\mathbf{j}}{T_2} - \frac{M_z(\mathbf{r},t) - M_0}{T_1}\mathbf{k}, \quad (1)$$

where $\mathbf{B}(\mathbf{r},t)$ is magnetic field at a spatial location $\mathbf{r}$, $M_0$、$T_1$、$T_2$ are tissue parameters. Due to the limitations of coil placement, in practical applications, the magnetization intensity $M_z$ in the z-direction is difficult to be directly detected. Therefore, our target is the transverse magnetization vector $M_\perp = \sqrt{M_x^2 + M_y^2}$. For the multi-spin echo sequence (Fig. 2) that measures the $T_2$ value of tissues, the transverse magnetization vector $M_\perp$ satisfies the Bloch equation that can be simplified as:

$$\frac{dM_\perp(\mathbf{r},t)}{dt} + \frac{M_\perp(\mathbf{r},t)}{T_2} = 0. \quad (2)$$

### B. PINN for $T_2$ Mapping

Traditional quantitative $T_2$ mapping for a single voxel (without variable $\mathbf{r}$) uses the least square (Fig. 1).

According to Bloch equation (2), the signal analysis formula can be obtained:

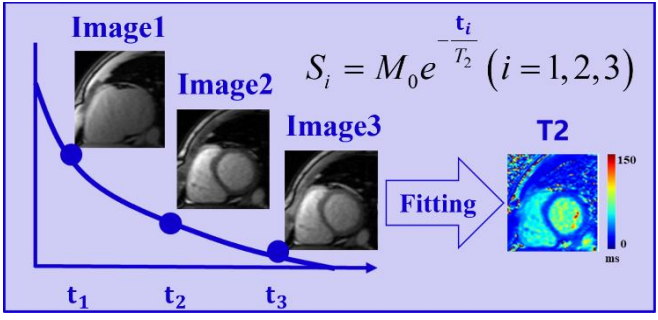

Fig. 1. Principle of least square for $T_2$ mapping.

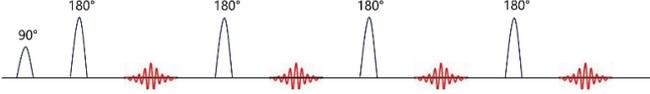

Fig. 2. Multi-spin echo sequence.

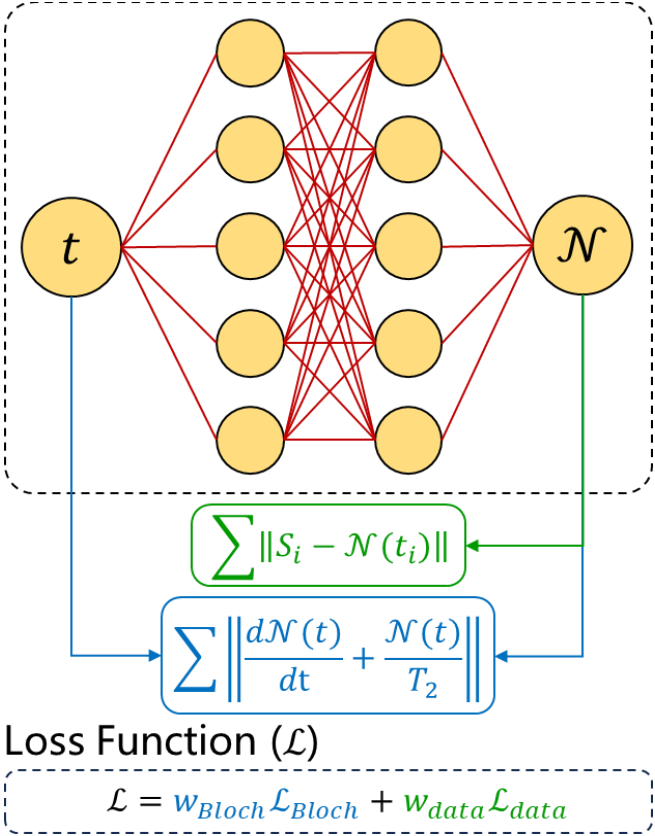

Fig. 3. Network structure of the Bloch equation-based PINN.

$$M_\perp(t) = M_0 e^{-\frac{t}{T_2}}, \quad (3)$$

where $M_0$ and $T_2$ are parameters to be quantified. For the measured data $M_\perp(t_i)$ in multiple time moments $t_i$, where $i \in \{1,2,\cdots,I\}$, $I$ is the number of different contrast images, we set $M_\perp(t_i)$ as $S_i$. According to the signal analysis formula, we can obtain $M_0$ and $T_2$ through fitting with least square.

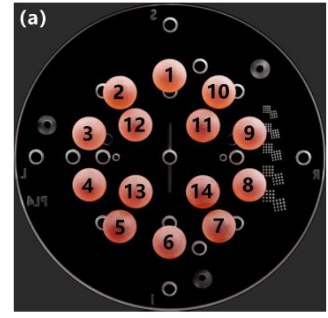

Phantom

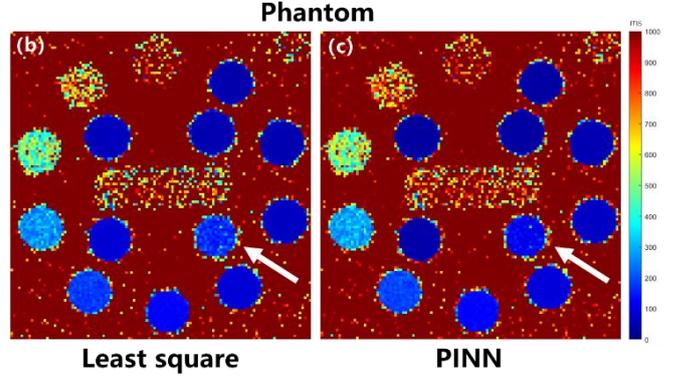

Least square    PINN

Fig. 4. $T_2$ mapping of $MnCl_2$ phantom. (a) An phtoto of the phantom, (b) and (c) are estimated $T_2$ values using the convetional least square and the proposed Bloch equation-based PINN, respectively.

However, for some sequences, the signal analysis formula cannot be obtained from Bloch equation, thus we use a nonlinear network is to approximate Bloch equation and obtain the quantitative parameters.

Here, we take multi-spin echo sequence as an example for quantitative $T_2$ mapping. The aim of PINN we propose is to approximate $M_\perp$ in Bloch equation (2) by network training. The structure of PINN is a fully connected network with two hidden layer (Fig. 3). The input of PINN is $t$, where $t$ is a set of K discrete time points in $T_2$ parameter interval, the network output is $\mathcal{N}(t)$. $\mathcal{N}(t)$ and $M_\perp(t)$ should satisfy the same equation:

$$\frac{d\mathcal{N}(t)}{dt} + \frac{\mathcal{N}(t)}{T_2} = 0. \quad (4)$$

The loss function of PINN is divided into two parts. The first part is a physics-informed loss for the Bloch equation (4):

$$\mathcal{L}_{Bloch} = \frac{1}{K}\sum_{k=1}^{K}\left\|\left(\frac{d\mathcal{N}(t)}{dt} + \frac{\mathcal{N}(t)}{T_2}\right)\bigg|_{t=t_k}\right\|, \quad (5)$$

where $\|\bullet\|$ is the $l_2$ norm, $k \in \{1,2,\cdots,K\}$. The second part is the loss between the network output and the measured realistic cardiac qMRI data as follows:

$$\mathcal{L}_{data} = \frac{1}{I}\sum_{i=1}^{I}\|S_i - \mathcal{N}(t_i)\|. \quad (6)$$

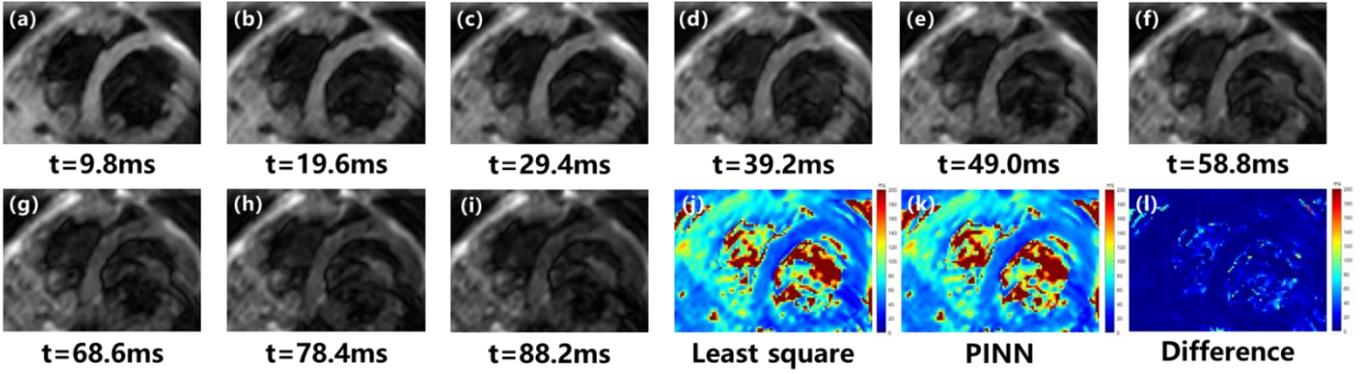

Fig. 5. $T_2$ mapping of cardiac MRI data from a helathy volunteer. (a)-(i) Nine different contrast images with different time t, (j) and (k) are estimated $T_2$ values using the convetional least square and the proposed Bloch equation-based PINN, respectively, (l) the difference between (j) and (k).

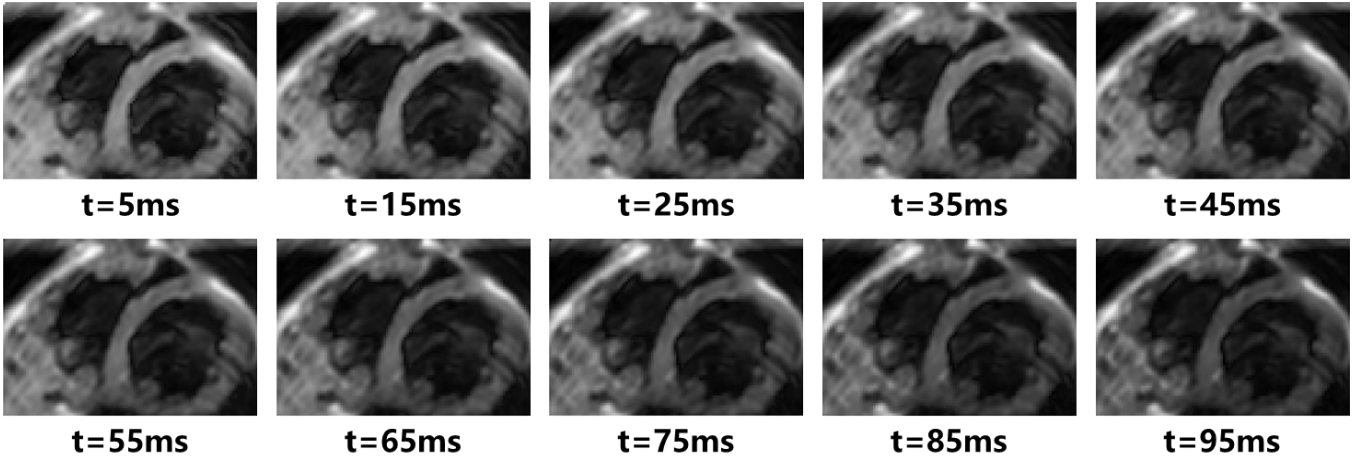

Fig. 6. Physically generated MRI data by usting the PINN at different time t.

Finally, the total loss function is a weighted sum of two parts:

$$\mathcal{L} = w_{Bloch}\mathcal{L}_{Bloch} + w_{data}\mathcal{L}_{data}, \quad (7)$$

where $w_{Bloch}$ and $w_{data}$ are two weights to balance importance of two terms.

## III. EXPERIMENTAL RESULT

### A. Phantom

The phantom consists of 14 tubes that are filled with different concentrations of $MnCl_2$ solution (Fig. 4(a)) [12]. We acquired the imaging data on an United Imaging Healthcare 3.0T MRI scanner. Imaging parameters include FOV = 240*240 mm$^2$ and slice thickness = 2 mm. PINN parameters are $K = 1001$, $C = 8$, $w_{Bloch} = 0.01$, and $w_{data} = 1$.

Figs. 4(b) and (c) show that the estimated $T_2$ values by two approaches are close to each other at most tubes. For the tube #14 that have small $T_2$ value (The standard $T_2$ is 5.592 ms [12]), the estimated $T_2$ values by PINN is closer to the reported standard one than the conventional least square approach.

### B. Healthy Volunteer

The cardiac imaging data were acquired from a healthy volunteer on a Philips 3.0T scanner. Imaging parameters are FOV = 300*300 mm$^2$ and slice thickness = 10 mm. PINN parameter include $K = 1001$, $C = 8$, $w_{Bloch} = 0.01$ and $w_{data} = 1$.

Given nine contrast images with different time $t$, the $T_2$ values estimated by two methods are comparable (Fig. 5).

### C. Data Generation

The PINN approximates the relationship from time $t$ to the transverse magnetization intensity $M_\perp(t)$ after training. Input time $t$, the corresponding transverse magnetization intensity $M_\perp(t)$ can be obtained for any voxel. Therefore, the trained PINN can be used to generate the MRI data at any time $t$ (Fig. 6). This may benefit physics-driven deep learning, particularly when a large amount of training data is not available [9, 13-15].

## IV. CONCLUSION

This paper proposes to incorporate the physical rule of MRI, the Bloch equation, into the neural network learning. No matter on phantom or realistic data, $T_2$ maps obtained for by two methods are comparable. Thus, the Bloch equation enables physics-informed neural network in parametric magnetic resonance imaging.

ACKNOWLEDGMENTS

This work was partially supported by the National Natural Science Foundation of China (62122064, 61971361, 62371410 and 62331021), the Natural Science Foundation of Fujian Province of China (2023J02005 and 2021J011184), the President Fund of Xiamen University (20720220063), and Nanqiang Outstanding Talent Program of Xiamen University. Thanks are due to Huajun She, Bei Liu and Wenlong Feng